\documentclass[fleqn,twoside]{article}
\usepackage{espcrc2}
\usepackage{graphicx}
\usepackage{times}

\title{Hidden Sector Models and Signatures}

\author{Zuowei Liu 
\address{C.N.\ Yang Institute for Theoretical Physics,
Stony Brook University,\\ Stony Brook, New York 11794-3840, USA}}

\begin{document}

\begin{abstract}
In the Stueckelberg extension of the Standard Model (StSM), matter 
in the hidden sector can act as dark matter. Due to an interplay of 
mixings produced by the usual Higgs mechanism and the Stueckelberg 
mechanism in the neutral gauge boson sector, the hidden sector matter 
acquires a milli charge. The Stueckelberg extension also produces a narrow 
width Z prime which is detectable at the Large Hadron Collider. 
The hidden sector dark matter naturally explains the PAMELA positron excess 
by means of a  Breit-Wigner enhancement through a Z prime resonance. We 
also discuss the origin of milli charge in the context of the kinetic mixing 
and the Stueckelberg mixing.
\vspace{-8cm}
\begin{flushright}{YITP-SB-09-28}\end{flushright}
\vspace{7cm}
\vspace{1pc}
\end{abstract}

\maketitle

\section{Introduction}

A Stueckelberg mechanism allows an Abelian gauge boson to develop a mass without 
the benefit of a Higgs mechanism (For the early history of the Stueckelberg mechanism 
see, \cite{stueckelberg:38,ogievetskii:62,Kalb:1974yc} 
and \cite{Cianfrani:2007tu,Cianfrani:2008sf}).
The Stueckelberg Lagrangian couples one Abelian gauge boson $A_{\mu}$ to one 
pseudo-scalar $\sigma$ in the following way 
\begin{equation}
{\cal L}= -\frac14 {F}_{\mu\nu}{F}^{\mu\nu} - \frac12 (m A_\mu + \partial_\mu
\sigma)(m A^\mu + \partial^\mu \sigma) 
\end{equation}
which is gauge invariant under the transformation 
\begin{eqnarray}
\delta A_{\mu} &=& \partial_\mu \lambda , \\
\delta \sigma &=& -m \lambda  .
\end{eqnarray}
The gauge boson mass growth through the Stueckelberg couplings can occur naturally in
compactifications of higher-dimensional string theory, supergravity, or even pure
gauge theory. Here we review the Stueckelberg  extension of the Standard 
Model. (For Stueckelberg extensions of the SUSY models, 
see \cite{Kors:2004ri,Kors:2004iz,Kors:2005uz,Feldman:2006wd,Feldman:2007nf,Feldman:2008en,Nath:2008ch,Feldman:2009wv}.)
We only review the Stueckelberg models where the extra $U(1)$ is not anomalous. 
(For the anomalous $U(1)$ case, see 
\cite{Anastasopoulos:2007qm,Kumar:2007zza,Anastasopoulos:2008jt,DeRydt:2008hw,Armillis:2008vp,Coriano:2008aw,Coriano:2008xa,Zhang:2008xq,Mambrini:2009ad}.)

\section{Stueckelberg Extension of the Standard Model (StSM)}
The Stueckelberg Lagrangian can extend the Standard Model 
to include additional $U(1)$ gauge fields \cite{Kors:2004dx}. 
The simplest case is to extend the Standard Model gauge group 
to have one additional $U(1)$, i.e., 
$SU(3)_C\times SU(2)_L \times U(1)_Y \times U(1)_X$. 
Thus in this case the total Lagrangian reads 
\begin{equation}
\begin{array}{rcl}
{\cal L_{\rm StSM}} &=& {\cal L_{\rm SM}}+{\cal L_{\rm St}},\\
{\cal L_{\rm St}} &= & -\frac14 {C}_{\mu\nu}{C}^{\mu\nu} 
+g_XC_{\mu}J_X^{\mu} \\
				&& - \frac12 (\partial_\mu\sigma+M_1 C_\mu + M_2 B_\mu )^2,
\end{array}
\label{eq:stsm}
\end{equation}
where $C_{\mu}$ is the gauge field associated with $U(1)_X$ and $B_\mu$ is the SM 
hypercharge gauge field; $g_X$ is the gauge coupling strength of the $U(1)_X$ and 
$J_X^{\mu}$ is the conserved fermion matter current that the gauge filed $C_\mu$ couples to. 
In general, the interaction between the gauge field $C_\mu$ and matter fields can include 
both the Standard Model fermion sectors (visible sector) and hidden fermion sector matter fields. 
However, the interaction between the $U(1)_X$ gauge boson and the visible sector matter can lead to 
modifications of the charge of the neutron which is measured to a high precision \cite{Kors:2005uz}. 
To avoid this potential violation, we assume that the Standard Model fields 
are neutral under $U(1)_X$.  
It is easy to check that the above Lagrangian is invariant under the $U(1)_X\times U(1)_Y$ 
transformation 
\begin{equation}
\delta_X (C_\mu,B_\mu,\sigma)=(\partial_\mu \lambda_X,0,-M_1\lambda_X),
\end{equation}
\begin{equation}
\delta_Y (C_\mu,B_\mu,\sigma)=(0,\partial_\mu \lambda_Y,-M_2\lambda_Y).
\end{equation}

The addition of the new gauge boson $C_\mu$ enlarges the vector boson 
neutral sector. In the basis $V_\mu^T=(C_\mu,B_\mu,A_\mu^3)$, 
the mass matrix then takes the following form
\begin{equation} 
M^2= \left[\matrix{
    M_1^2             & M_1^2\epsilon                                & 0 \cr
    M_1^2\epsilon         &M_1^2\epsilon^2 + \frac{1}{4}v^2g_Y^2  & -\frac{1}{4}v^2g_2g_Y \cr
      0                   & -\frac{1}{4}v^2g_2g_Y            & \frac{1}{4}v^2g_2^2
}\right],
\label{mass}
 \end{equation}
where $\epsilon  \equiv M_2/M_1$; 
$g_2$ and $g_Y$ are the $SU(2)_L$ and $U(1)_Y$ gauge coupling
constants, and are normalized so that $M_W^2=g_2^2v^2/4$. 
This allows one to choose $\epsilon$ and $M_1$ as two independent
parameters to characterize physics beyond the SM. 
It is easily checked that  $\det(M^2)=0$ which implies that one
of the eigenvalues is zero, whose eigenvector we identify with the
photon. The remaining two eigenvalues are non-vanishing and
correspond to the $Z $ and $Z'$ bosons.

The symmetric matrix $M^2$ can be diagonalized by an 
orthogonal transformation, $V={\cal O}E$, 
with $E_{\mu}^T = ( Z'_\mu, Z_\mu, A_\mu^\gamma)$. 
The transformation matrix ${\cal O}$ can be 
parameterized by three angles as follows
\begin{eqnarray}
{\cal O}= \left[\matrix{ 
c_\psi c_\phi -s_\theta s_\phi s_\psi & s_\psi c_\phi +s_\theta s_\phi c_\psi & - c_\theta s_\phi \cr 
c_\psi s_\phi +s_\theta c_\phi s_\psi & s_\psi s_\phi-s_\theta c_\phi c_\psi & c_\theta c_\phi\cr
-c_\theta s_\psi &  c_\theta c_\psi & s_\theta }\right]
\end{eqnarray}
where $s_\theta=\sin\theta$, $c_\theta=\cos\theta$, etc. 
The angels are defined as \cite{Kors:2005uz}
\begin{equation}
\tan\phi =  \epsilon,
\end{equation}
\begin{equation}
\tan\theta =\frac{g_Y}{g_2\sqrt{1+\epsilon^2}},
\end{equation}
\begin{equation}
\tan \psi = \frac{\tan\theta\tan\phi M^2_W}
{\cos\theta(M^2_{Z'}-M^2_W(1+\tan^2\theta))}.
\end{equation}
There is also  a modification of the expression of the electric charge in terms of the 
fundamental physics couplings. Thus if we write the EM interaction in the form $e
A_\mu^\gamma J^\mu_{\rm em}$ the expression for $e$ is given by
\begin{equation}
\frac{1}{e^2} = \frac{1}{g_2^2}+\frac{1+\epsilon^2}{g_Y^2}.
\label{electric}
\end{equation}
Thus $g_Y$ is related to $g_Y^{\rm SM}$ by 
\begin{equation}
g_Y=g_Y^{\rm SM}\sqrt{1+\epsilon^2}.
\label{gy}
\end{equation}

In the limit $\epsilon \to 0$, when the Abelian $C_\mu$ field decouples from 
SM, the mixing angles $\phi$ and $\psi$ vanish, and the hypercharge coupling 
$g_Y$ recovers its SM value.

\section{Stueckelberg Extension of the Standard Model with Kinetic Mixing (StkSM)}

We discuss here the StSM model in the context of kinetic mixing between $U(1)_X$ 
and $U(1)_Y$. Such kinetic mixing is quite generic, and can arise in a variety of ways 
in a broad class of models. We will characterize the kinetic mixing term 
as follows 
\begin{equation}
{\cal L_{\rm KM}}=-\frac{\delta}{2}C_{\mu\nu}B^{\mu\nu}, 
\end{equation}
and the total Lagrangian of StkSM is 
${\cal L_{\rm StkSM}}={\cal L_{\rm StSM}}+{\cal L_{\rm KM}}$ 
where ${\cal L_{\rm StSM}}$is given in Eq.(\ref{eq:stsm}). 
Thus, the model consists of both the kinetic mixing and the Stueckelberg 
mass mixing \cite{Feldman:2007wj,Abel:2008ai}. 
The kinetic mixing Lagrangian leads to non-diagonal elements in the kinetic 
energy Lagrangian in the basis $V_\mu^T=(C_\mu,B_\mu,A_\mu^3)$
\begin{eqnarray}
{\cal K}= \left[\matrix{ 
1 & \delta & 0 \cr  
\delta & 1 & 0 \cr 
0 & 0 & 1}\right].
\label{km}
\end{eqnarray}
A simultaneous diagonalization of both the kinetic energy terms, Eq.(\ref{km}), and the 
mass matrix, Eq.(\ref{mass}) can be achieved by a transformation $T=KR$, which 
is a combination of a $GL(3)$ transformation ($K$) and an orthogonal transformation 
($R$). This allows one to work in the mass basis with the transformation 
$E_\mu=R^TK^TV_\mu$. A special choice of the $GL(3)$ transformation ($K$) as follows 
\begin{eqnarray}
{K}= \left[\matrix{ 
1 & -\delta/\sqrt{1-\delta^2} & 0 \cr  
0 & 1/\sqrt{1-\delta^2}& 0 \cr 
0 & 0 & 1}\right]
\label{gl3}
\end{eqnarray}
diagonalizes the kinetic energy terms, and at the same time transforms the mass matrix, 
Eq.(\ref{mass}) into 
\begin{equation} 
M^2= \left[\matrix{
    M_1^2             & M_1^2\bar\epsilon                                & 0 \cr
    M_1^2\bar\epsilon         &M_1^2\bar\epsilon^2 + \frac{1}{4}v^2\bar g_Y^2  
    & -\frac{1}{4}v^2g_2 \bar g_Y \cr
      0                   & -\frac{1}{4}v^2g_2 \bar g_Y            & \frac{1}{4}v^2g_2^2
}\right]. 
\label{mass2}
\end{equation}
The above mass matrix in StkSM looks exactly the same as the one in StSM, Eq.(\ref{mass}), 
except that the parameters $\epsilon$ and $g_Y$ get redefined as follows
\begin{equation}
\bar\epsilon = \frac{\epsilon - \delta}{\sqrt{1-\delta^2}},
\end{equation}
 \begin{equation}
\bar g_Y = \frac{g_Y}{\sqrt{1-\delta^2}}.
\end{equation}
Thus the orthogonal transformation $R(\bar\epsilon, \bar g_Y)$ has the same form 
as the orthogonal transformation ${\cal O}(\epsilon, g_Y)$ in StSM. 
The neutral sector interaction can now be written in the form 
\begin{equation}
{\cal L}_{\rm NC}=J^{\mu T} K(\delta) R(\bar\epsilon, \bar g_Y) E_\mu
\end{equation}
where $J^{\mu T}=(g_XJ_X^\mu,g_YJ_Y^\mu,g_2J_2^{3\mu})$, so one finds that  
\begin{eqnarray}
{J^{\mu T} K(\delta)}= \left[\matrix{ 
g_XJ_X^\mu \cr
K_{12}g_XJ_X^\mu  + \bar g_YJ_Y^\mu  \cr
g_2J_2^{3\mu}
}\right]^T
\label{nc}
\end{eqnarray}
where $K_{12}=-\delta/\sqrt{1-\delta^2}$ is the matrix element of 
the $GL(3)$ transformation ($K$), and we have used $\bar g_Y$ to absorb 
the matrix element $K_{22}$.  
If one restricts $J_X$ to only include the hidden sector matter, 
the interactions between Standard Model fermions and $E_\mu$ depend on 
$\bar \epsilon$ and $\bar g_Y$, not explicitly on $\delta$, and have the same 
form as in StSM. So one expects the same electric charge modification 
as in Eqs.(\ref{electric}, \ref{gy}) in StkSM
\begin{equation}
\bar g_Y=g_Y^{\rm SM}\sqrt{1+\bar\epsilon^2}. 
\label{gy2}
\end{equation}
When substituting the modifications of the hypercharge coupling constants 
Eq.(\ref{gy}) and Eq.(\ref{gy2}) back into the mass matrices Eq.(\ref{mass}) 
and Eq.(\ref{mass2}), one finds that the mass matrices are identical except for the 
change of variable $\epsilon \to \bar \epsilon$. Thus the orthogonal transformation 
$R(\bar\epsilon)$ depends only on $\bar \epsilon$, and has the form of  
the orthogonal transformation ${\cal O}(\epsilon)$ in StSM, i.e., $R={\cal O}$. 
So one can conclude that 
the neutral sector interactions in StkSM are the same as in StSM, in the absence of 
the hidden sector matter. In order to discriminate the StkSM from StSM, one has 
to include the hidden sector.

The hidden sector matter is milli-charged with respect to the photon 
due to the small mixing between the two $U(1)$ fields. In StkSM, 
the milli charge ($Q_{\rm milli}$) of the  hidden sector matter 
can be expressed as 
\begin{equation}
Q_{\rm milli} \propto R_{13}(\bar\epsilon)
+K_{12}(\delta)R_{23}(\bar\epsilon).
\end{equation}
It is interesting to note that the milli charge ($Q_{\rm milli}$) vanishes 
when taking the limit $\epsilon \to 0$. Thus in Stueckelberg 
models, it is the Stueckelberg mass mixing $\epsilon$ that generates the 
milli charge of hidden sector matter, not the kinetic mixing $\delta$. 
We further discuss this topic in more details in the next section.

\section{Milli Charge in Stueckelberg models}

We illustrate the mechanism which generates the milli charge in Stueckelberg models.  
Consider a kinetic mixing model with two gauge fields $A_{1\mu}, A_{2\mu}$ 
corresponding to the gauge groups $U(1)$ and $U(1)'$. 
We choose  the following Lagrangian
$\mathcal{L} =\mathcal{L}_0 + \mathcal{L}_1$ where
\begin{eqnarray}
\mathcal{L}_0 &=&
    - \frac{1}{4}F_{1\mu\nu}F_1^{\mu\nu}
    - \frac{1}{4}F_{2\mu\nu}F_2^{\mu\nu}
    - \frac{\delta}{2}F_{1\mu\nu}F_2^{\mu\nu},\nonumber\\
 \mathcal{L}_1 &=&
     J'_{\mu}A_1^{\mu}
    +J_{\mu}A_2^{\mu}.
       \label{onlykt}
\end{eqnarray}
Here $J_\mu$ is the physical source arising from quarks and leptons,  
and $J'_\mu$ is the source from the hidden sector. 
To put  the kinetic energy term in its canonical form, one may use the transformation
\begin{equation}
 K_0=\left[\matrix{1/\sqrt{1-\delta^2} & 0 \cr
-\delta/\sqrt{1-\delta^2} &1}\right].
 \end{equation}
However, the transformation is not unique, since $K=K_0 R$ instead of $K_0$ 
would do as well, where $R$ is an orthogonal matrix
\begin{equation}
    R=\left[\matrix{ \cos\theta & -\sin\theta \cr \sin\theta &
    \cos\theta}\right].
\end{equation}
Thus  $\mathcal{L}_1$ can be rewritten in terms of the new basis 
$A$ and $A'$
\begin{eqnarray}
 \mathcal{L}_1 =
    A'^{\mu}  \left[\frac{\cos\theta}{\sqrt{1-\delta^2}} J'_{\mu} +  
    \left(\sin\theta-\frac{\cos\theta\delta}{\sqrt{1-\delta^2}}\right) J_{\mu} \right]
    \nonumber\\ + A^{\mu}  \left[  -\frac{\sin\theta}{\sqrt{1-\delta^2}} J'_{\mu}  +
     \left( \cos\theta+\frac{\sin\theta\delta}{\sqrt{1-\delta^2}}\right)J_{\mu}
     \right].
\end{eqnarray}
In this case we see that both $A$ and $A'$ can interact with 
$J$ and $J'$. However,  one may choose $\theta$ to get asymmetric solutions. 
For instance, one can take $\theta =\arctan\left[\delta/\sqrt{1-\delta^2}\right]$ 
to decouple $A'$ from $J$. Here both $A$ and $A'$ are massless.

Next we consider a model with kinetic mixing where a Stueckelberg mechanism
generates a mass term of the type 
\begin{eqnarray}
{\cal{L}}_{\rm{Mass}} = -\frac{1}{2} M_1^2  A_{1\mu}A_1^{\mu}
-\frac{1}{2} M_2^2 A_{2\mu}A_2^{\mu} \\\nonumber
- M_1M_2 A_{1\mu}A_2^{\mu}.
\end{eqnarray} 
In this case diagonalizaton of  the mass  matrix fixes $\theta$ so that
\begin{equation}
\theta=\arctan\left[\frac{\epsilon\sqrt{1-\delta^2}}{1-\delta\epsilon}\right],
\end{equation}
where $\epsilon \equiv M_2/M_1$ is the Stueckelberg mass mixing parameter. 
Thus the interaction Lagrangian in the new basis is given by
\begin{eqnarray}
{\cal{L}}_{1} = \frac{ (\epsilon-\delta) J_{\mu}+(1-\delta\epsilon) J_{\mu}' }
 {\sqrt{1-2\delta\epsilon+\epsilon^2}\sqrt{1-\delta^2}} A_M^\mu \\\nonumber
 + \frac{J_{\mu}- \epsilon J_{\mu}' }
 {\sqrt{1-2\delta\epsilon+\epsilon^2}} A^{\mu}_{\gamma}.
\end{eqnarray}
Here for the case  $\epsilon =0$ one finds that the massless state, 
the photon $A_{\gamma}^{\mu}$, no longer couples with the 
hidden sector. We  conclude, therefore, that in the absence of the
Stueckelberg mass mixing, for the case when only one mode is
massless, there are no milli-charged particles coupled to the photon
field. Thus milli-charge couplings appear in this case only when
the Stueckelberg mixing parameter $\epsilon$ is introduced. For 
the case when only one mode is massless the kinetic mixing by itself 
does not allow for a milli charge. 

Thus, the milli charge in Stueckelberg model 
is to be contrasted with the model where one has two massless modes 
(the photon and the paraphoton) \cite{Holdom:1985ag} 
and the photon can couple with the hidden sector because of 
kinetic mixing generating milli-charged couplings. 
(An analysis with kinetic mixing and mass mixings 
of a different type is also considered in \cite{Holdom:1990xp}).

\begin{table}[thb]
\hspace{-0.7cm}
\begin{tabular}{lrrrr}
\hline
Exp ($\Delta$)     &LEP Fit(Pull)          &St Fit(Pull)      \\
\hline
$\Gamma_Z$=2.4952(.0023)      			&2.4956(-0.17)       	&2.4956(-0.17)     \\
$\sigma_{\rm had}$=41.541(.037)       	&41.476(1.76)       	&41.469(1.95)    \\
$R_e$=20.804(.050)        				&20.744(1.20)       	&20.750(1.08)     \\
$R_\mu$=20.785(.033       				&20.745(1.21)      	&20.750(1.06)     \\
$R_\tau$=20.764(.045)       				&20.792(-0.62)       	&20.796(-0.71)    \\
$R_b$=0.21643(.00072)     				&0.21583(0.83)     	&0.21576(0.93)    \\
$R_c$=0.1686(.0047)       				&0.17225(-0.78)      	& 0.17111(-0.53)   \\
$A^{(0,e)}_{FB}$=0.0145(.0025)       		&0.01627(-0.71)      &0.01633(-0.73)   \\
$A^{(0,\mu)}_{FB}$=0.0169(.0013)      	&0.01627(0.48)      & 0.01633(0.44)   \\
$A^{(0,\tau)}_{FB}$=0.0188(.0017)     	&0.01627(1.49)      & 0.01633(1.45)   \\
$A^{(0,b)}_{FB}$=0.0991(.0016)      		&0.10324(-2.59)      & 0.10344(-2.71)   \\
$A^{(0,c)}_{FB}$=0.0708(.0035)      		&0.07378(-0.85)      &0.07394(-0.90)   \\
$A^{(0,s)}_{FB}$=0.098(.011)        		&0.10335(-0.49)      &0.10355(-0.50)   \\
$A_e$=0.1515(.0019)       				&0.1473(2.21)        &0.1476(2.05)         \\
$A_\mu$=0.142(.015)        				&0.1473(-0.35)        &0.1476(-0.37)        \\
$A_\tau$=0.143(.004)        				&0.1473(-1.08)        &0.1476(-1.15)        \\
$A_b$=0.923(.020)        					&0.93462(-0.58)      &0.93464(-0.58)       \\
$A_c$=0.671(.027)        					&0.66798(0.11)      &0.66812(0.11)         \\
$A_s$=0.895(.091)         					&0.93569(-0.45)      &0.93571(-0.45)       \\
\hline 
                                                                       & ($\chi^2$=25.0)      &($\chi^2$=25.2)			\\
\hline
\end{tabular}
\caption{Fits to 19 $Z$ pole observables. The 1st column is given by the PDG \cite{pdg} 
with uncertainties $\Delta$. The 2nd column is from the SM Fit of the LEP EWWG \cite{:2005ema}. 
The 3rd column is an analysis for one StkSM model with inputs $\epsilon =0.06$, $\delta= 0.03$, 
and $M_1 = 200$ GeV. The Pull is defined as $(\rm Exp  - \rm Fit)/\Delta$,  
and $\chi^2 = \sum{\small{\rm{Pull}}^2}$. Taken from \cite{Feldman:2007wj}. 
}\label{tab:fit}
 \end{table}

\section{Electroweak Constraints}

Since the Standard Model is very successful in explaining physics at the 
electroweak scale, effects from new physics receive stringent constraints 
from electroweak observables. We concentrate our analysis on the StkSM 
where the effective coupling of new physics is 
\begin{equation}
\bar\epsilon = \frac{\epsilon - \delta}{\sqrt{1-\delta^2}}. 
\end{equation}
As discussed earlier, StkSM and StSM are indistinguishable when considering 
electroweak constraints for the case where the hidden sector is absent. For the 
analysis of electroweak constraints, we do not consider the hidden sector matter.

The Stueckelberg mechanism predicts a mass shift for vector $Z$ boson. To be 
consistent with electroweak precision tests, such a shift must be small enough so that 
it lies within the error corridor of SM prediction. In the limit of $M_1\gg M_Z$, 
one finds an approximate upper bound on the mixing parameter $\bar\epsilon$ \cite{Feldman:2006ce}
\begin{equation}
|\bar\epsilon|\le 0.061\sqrt{1-(M_Z/M_1)^2}.
\end{equation}

Another constraint comes from the LEPII experiments where the new physics 
is typically characterized by the contact interaction parameter 
$\Lambda$. Among different contact interactions analyzed by LEPII group, 
one finds that the $\Lambda_{VV}> (21.7, 17.1)$ TeV \cite{Leptwo} 
usually gives the most stringent bound on $\bar\epsilon$. 
The StkSM predicts the theoretical value of $\Lambda_{VV}$ through the following formula 
\begin{equation}
\Lambda_{VV}=\frac{M_{Z'}}{M_Z}\sqrt{\frac{4\pi}{\sqrt{2}G_Fv_e^{'2}}}.
\end{equation}

\begin{figure}[t]
\hspace{-0.6cm}
\includegraphics[width=8cm,height=8cm]{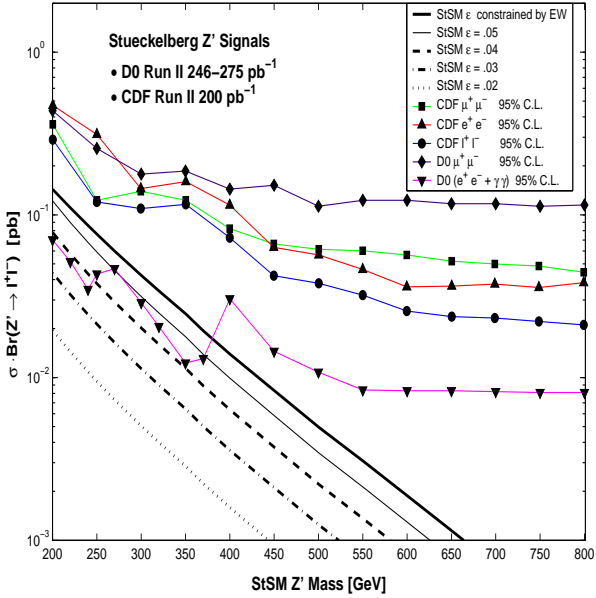}
\caption{  
An exhibition of the $Z'$ leptonic signal for different 
$\epsilon$ values in StSM, in comparison with  
the CDF \cite{cdfdata} data with 200 pb$^{-1}$ 
and D0 \cite{Abazov:2005pi} data with 246-275 pb$^{-1}$. 
The data (at 95\% C.L.) puts a lower limit 
of about  250 GeV on $M_{Z'}$ for  $\epsilon \approx 0.035$ 
and 375 GeV for $\epsilon \approx 0.06$. 
Figure taken from \cite{Feldman:2006ce}.
}
\label{fig:cdf1}
\end{figure}

However, even more stringent constraints come from fits to the high 
precision LEP data on the branching ratios of the $Z$ decay and 
the various asymmetries at the $Z$ pole, when one demands that
the $\chi^2$ of the StkSM fits lie within 1\% of that of the Standard 
Model. Table.(\ref{tab:fit}) gives one example of the global fits to 19 
observables for one specific StkSM model in the parameter space with 
$\bar\epsilon\sim 0.03$ and $M_{Z'}\sim 200$ GeV. 
As exhibited in Table.(\ref{tab:fit}), the StkSM fits the 
19 electroweak observables as well as the Standard 
Model does. As we will see in the next section, at 200 GeV scale, a $Z'$ 
signal controlled by the effective coupling $\bar\epsilon$ can be probed 
at the Tevatron collider with current and near future integrated luminosity. 
For a $Z'$ resonance at a higher scale, one expects the constraints from 
the electroweak observable fits to become more relaxed than that at the 
scale close to the $Z$ scale.

\section{Probing Narrow $Z'$ Resonance at Tevatron}

One of the most promising channels to discover a new resonance at hadron colliders 
is the clean leptonic final states in the Drell-Yan process. In StSM, we are more interested 
in the energy region where two partons fuse into a $Z'$ boson and then the $Z'$ boson 
decays into dileptons.

\begin{figure}[t]
\hspace{-0.6cm}
\includegraphics[width=8cm,height=8cm]{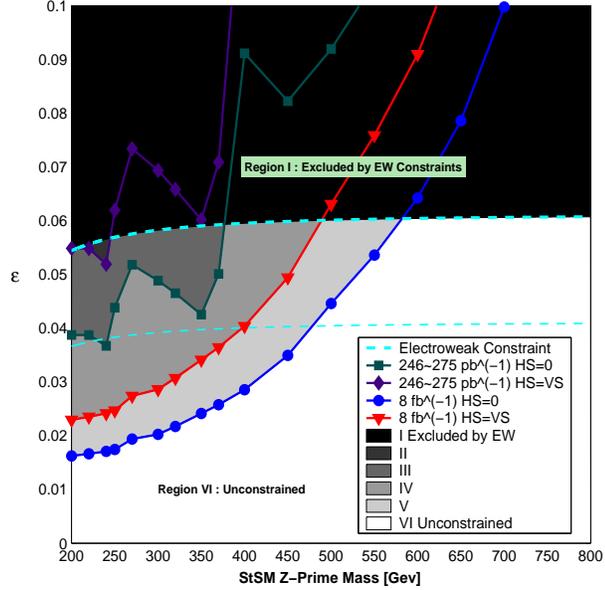}
\caption{
Exclusion plots in the $\epsilon -M_{Z'}$ plane utilizing
the more sensitive D0  \cite{Abazov:2005pi}  
$e^+e^-+\gamma\gamma$ mode with 
(a) the  246-275 pb$^{-1}$ of data, and 
(b) 8 fb$^{-1}$ of data  where an extrapolation of the
sensitivity curve is used. 
Cases with a naive inclusion of hidden sector decay are also shown.  
Regions  II, III, IV, and V are constrained by the conditions given at their
respective boundaries. Figure taken from \cite{Feldman:2006ce}.
} 
\label{fig:cdf2}
\end{figure}

A detailed analysis of the Drell-Yan cross section for the process 
$p\bar p\to Z' \to \ell^+ \ell^-$ as a function of  $M_{Z'}$ 
in StSM is given in Fig.(\ref{fig:cdf1}).
The analysis is done at $\sqrt s = 1.96$ TeV, 
with a flat $K$ factor of $1.3$ for the appropriate 
comparisons with other models and with the CDF \cite{cdfdata} 
and D0 \cite{Abazov:2005pi} combined data in the dilepton channel. 
(For more recent CDF data, see \cite{Aaltonen:2008vx}.)  
We integrate the dileptonic differential cross section near the $Z'$ resonance, 
and compare it with the experimental 95\% C.L.\ limits. 
Combining the theoretic calculation of the $Z'$ resonance and the 
Tevatron data, one finds that the Stueckelberg $Z'$ for the case
$\epsilon \approx 0.06$  is eliminated up to about 375 GeV 
(at 95\% C.L.). This lower limit decreases as 
 $\epsilon$ decreases  but the data  still constrains the model
 up to $\epsilon\approx 0.035$.

\begin{figure}[t]
\hspace{-0.6cm}
\includegraphics[width=8cm,height=7cm]{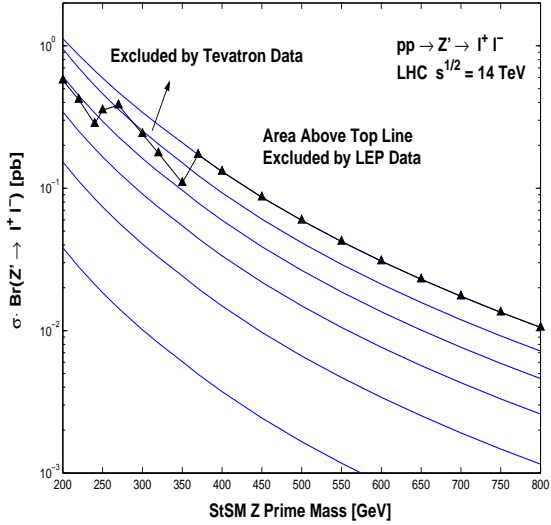}
\caption{
The prediction of the production cross section $ \sigma(pp \to Z') \cdot 
Br(Z' \to \ell^+\ell^-) $ [pb] for different $Z'$ masses in the StSM at the LHC with
the inclusion of the Tevatron constraints. The curves in
descending order  correspond  to values of $\epsilon$ from .06 to
.01 in steps of .01. Figure taken from \cite{Feldman:2006wb}.
}
\label{fig:lhc1}
\end{figure}

In Fig.(\ref{fig:cdf2}) we give the exclusion plots 
in the $\epsilon -M_{Z'}$ plane using the 246-275 
pb$^{-1}$ of data from D0 \cite{Abazov:2005pi} 
and also using the total integrated luminosity of 8 fb$^{-1}$ 
expected at the Tevatron. An analysis including the hidden sector with 
$\Gamma_{\rm HS}=\Gamma_{\rm VS}$ is also exhibited. 
The exclusion plots show that even the hidden sector is beginning 
to be constrained and  these constraints will become even more 
severe with future data. The inclusion of the hidden sector here 
is only for illustrative purposes. A formal treatment of the hidden 
sector matter will be discussed later in the context of dark matter and 
LHC signatures.

\section{LHC Signatures of Narrow $Z'$ Resonance}

Here we give an analysis for the exploration of  the $Z'$  boson
at the LHC. The models considered here only consist of the case 
when the hidden sector fermions are absent or are kinematically 
inaccessible. If the $Z'$ boson can decay into the hidden sector matter, 
the invisible decay widths usually dominate the visible widths, and thus dilute 
the LHC signatures. We discuss this further in details in the context of 
dark matter in the next section.

\begin{figure}[t]
\hspace{-0.6cm}
\includegraphics[width=8cm,height=7cm]{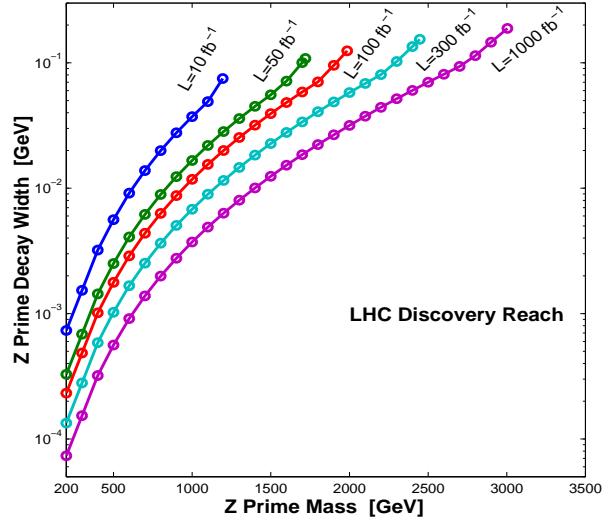}
\caption{
An analysis of the discovery reach of the LHC for StSM 
$Z'$ decay widths and $Z'$ masses.  
The regions that can be probed at LHC for a given luminosity 
are to the left and above each curve. 
Figure taken from \cite{Feldman:2006wb}.} 
\label{fig:lhc2}
\end{figure}

At the LHC, the promising signature channel for searching 
the $Z'$ resonance is also the di-lepton final state, 
$pp\to Z'\to \ell^+\ell^-$. The analysis of 
Fig.(\ref{fig:lhc1}) predicts the dilepton signal 
$\sigma(pp\to Z') \cdot Br(Z'\to \ell^+\ell^-)$ in StSM  
that can be detected at LHC. 
Also plotted here is the signature region that has already been probed by 
Tevatron data in the mass regions up to 800 GeV.

Further we investigate the capability of the LHC in probing the mass and 
decay width of the Stueckelberg $Z'$ boson. The criteria for the discovery  
of new physics is taken as $S>{\rm Max}(5\sqrt {\rm SM}, 10)$. 
In Fig.(\ref{fig:lhc2}) we give the discovery reach for finding the 
StSM $Z'$ for different decay widths as a function of the $Z'$ 
mass for luminosity in the range 10 fb$^{-1}$ and 1000 fb$^{-1}$. 
In the analysis  we have assumed that detector effects can lead to 
signal and background losses of 50 percent. Here one finds that 
the LHC can probe a 100 MeV $Z'$ up to about 2.75 TeV 
and a 10 MeV width  up to a $Z'$ mass of about  1.5 TeV.

\begin{figure}[t]
\hspace{-0.6cm}
\includegraphics[width=8 cm,height=7cm]{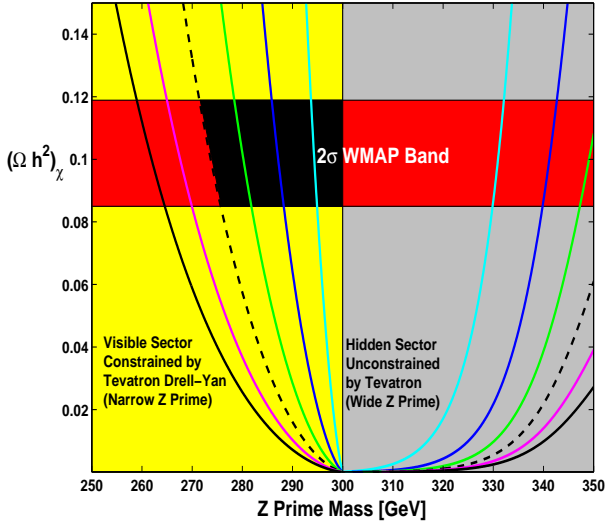}
\caption{
An analysis of  the relic density of milli-charged particles 
arising in the StSM from the hidden sector for 
the case $M_{\chi} = $ 150 GeV, $\epsilon$ =(.01-.06) 
with .01 for the innermost curve and moving outward in 
steps of .01. The (yellow, grey) regions ($M_{Z'}<2M_{\chi}$,
$M_{Z'}>2M_{\chi}$) correspond to a (narrow, broad) 
$Z'$ resonance, and the WMAP 3-year relic density 
constraints are satisfied for  both cases. 
Figure taken from \cite{Feldman:2007wj}.}
\label{fig:dark1}
\end{figure}

\begin{figure}[htb]
\hspace{-0.6cm}
\includegraphics[width=8 cm,height=7cm]{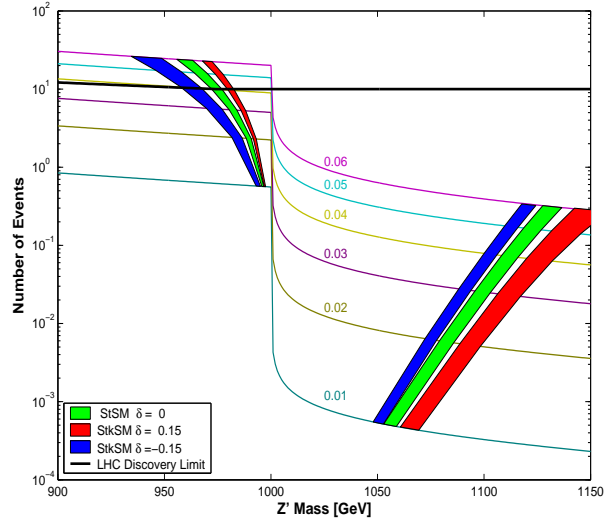}
\caption{ An exhibition of the dilepton signal as given by the
number of events for 10 fb$^{-1}$ of integrated luminosity at 
the LHC consistent with the 3 year WMAP \cite{Spergel:2006hy} 
relic density constraint  as a function of  $M_{Z'}$ when 
$M_{\chi}=500$ GeV. The curves in ascending order are 
for values of $\bar{\epsilon}$ in the range $(0.01-0.06)$ in 
steps of 0.01. Figure taken from \cite{Feldman:2007wj}.} 
\label{fig:dark2}
\end{figure}

\section{Milli Charge Dark Matter}

The milli-charged hidden sector matter in the Stueckelberg model 
is a natural candidate for the dark matter in the universe. 
Consider the model in which the hidden sector matter is 
a Dirac fermion $\chi$ which couples to $U(1)_X$ field via 
\cite{Cheung:2007ut,Cheung:2007uu,Feldman:2007wj}
\begin{equation}
{\cal L_{\rm int}} =g_XQ_X
\bar\chi \gamma^\mu \chi C_\mu
\end{equation}
where $Q_X$ is the quantum number. Thus, the interactions between 
the hidden Dirac fermions and Standard Model particles are connected 
by vector bosons through the weak coupling parameter $\bar\epsilon$
\begin{equation}
\chi\bar\chi \leftrightarrow 
A_{\gamma}/Z/Z' \leftrightarrow f \bar f. 
\end{equation}
Because the Dirac fermion $\chi$ has suppressed coupling to $A_{\gamma}$ and $Z$, 
and because $Z'$  has suppressed coupling to the Standard Model fermions, $\chi$ will always 
couple weakly to Standard Model fermions. Thus it usually requires 
the Dirac fermion to annihilate near the vicinity of the $Z$ or $Z'$ pole 
to be consistent with the relic abundance measured in WMAP experiments.

We discuss here only the annihilation near the $Z'$ resonance. To compute 
the relic density of the universe, one has to properly integrate the annihilation 
cross section for various $\chi$ particles with different kinetic energy, 
and then obtain the thermally averaged annihilation rate 
$\langle \sigma v \rangle$. The thermal average near a resonance 
can produce significant deviations from naive estimates for the calculation 
of the dark matter relic density.

An example of the relic density analysis in StSM where 
$M_{\chi} = $ 150 GeV, $\epsilon$ =(.01-.06) is exhibited in 
Fig.(\ref{fig:dark1}). It is shown that the relic density constraints 
from the 3 year WMAP data \cite{Spergel:2006hy} 
can be satisfied over a broad range of the parameter 
space on both sides of the $Z'$ pole.

As discussed earlier, one can discriminate the StkSM from StSM 
via the hidden sector interaction. We illustrate this effect in 
Fig.(\ref{fig:dark2}) along with LHC signatures. One finds that 
different bands that satisfy the relic density with different 
kinetic mixing parameter $\delta$ occupy different regions in the 
LHC signature space (the widths of the bands 
are translated from the error corridor of the relic density).

Another remarkable phenomenon from the hidden sector is its influence on 
the discovery reach of the $Z'$ resonance at LHC.  
When $2M_\chi<M_{Z'}$, $Z'$ can decay into hidden sector matter. 
The hidden sector decay width usually dominates the visible sector decay widths  
due to the normal coupling strength between $\chi$ and $Z'$. 
In this case, the $Z'$ produced at LHC will decay predominantly into 
hidden sector, and thus the hidden sector decay dilutes the visible 
leptonic signals at LHC as shown in Fig.(\ref{fig:dark2}). 
We note in passing that the searches of milli charged massive particles at 
the LHC are extremely difficult, and thus rely on sophisticated search strategies  
(for instance, see \cite{Chen:2009gu}). 
When $2M_\chi>M_{Z'}$, the hidden 
sector decay is turned off, the dileptonic signature is still the promising 
discovery channel at the LHC as discussed earlier.

\begin{figure}[t]
\includegraphics[width=8cm, height=7cm]{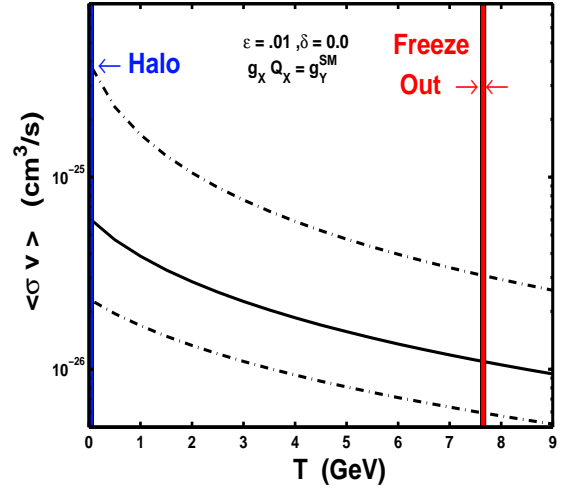}
\caption{ 
An exhibition of the  dependence of $\langle \sigma v\rangle$ 
on temperature for Stueckelberg models  
with $M_\chi =(150,151.5,153)$ GeV and $M_{Z'}= 298$ GeV. 
The annihilation near a pole generates a significant enhancement of 
$\langle \sigma v\rangle_{H}$ in the halo relative to 
$\langle \sigma v\rangle {X_f}$ at the freezeout. 
The natural Breit-Wigner enhancement of 
$\langle \sigma v\rangle_H$ obviates the 
necessity of using very large boost factors.
Figure taken from \cite{Feldman:2008xs}.}
\label{fig:bw}
\end{figure}

\begin{figure}[thb]
\includegraphics[width=8cm, height=7cm]{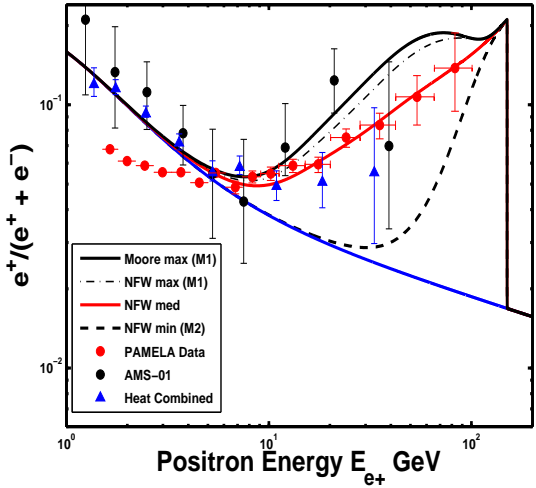}
\caption{ 
Positron spectrum including the monochromatic source 
and continuum flux in StSM with model inputs 
$\epsilon= 0.006$, $M_{Z'}= 298$ GeV and $M_\chi =150$ GeV, 
for various halo profiles and diffusion models. 
Also plotted here is the PAMELA  data \cite{Adriani:2008zr},
along with the AMS-01 and HEAT data 
\cite{Barwick:1997igAguilar:2007yf}. 
The background flux ratio is the 
decaying solid (blue) lower curve. 
Figure taken from \cite{Feldman:2008xs}.}
\label{fig:pamela}
\end{figure}

\section{PAMELA Positron Excess and Breit-Wigner Enhancement}

The recent positron excess observed by PAMELA satellite 
\cite{Adriani:2008zr} can be explained by dark matter 
annihilation in the galactic halo among other possible 
solutions. The positron excess requires the dark matter annihilation 
cross section in the galactic halo to be much larger 
than the annihilation cross section that gives rise to 
the relic density. This often leads to large boost factors 
whose physical origins are uncertain. However, this 
difficulty can be overcome with annihilations near a 
$Z'$ pole which is a narrow resonance  
in Stueckelberg models \cite{Feldman:2008xs}.

An illustration of the enhancement near the Breit-Wigner $Z'$ pole 
is given in Fig.(\ref{fig:bw}). The Breit-Wigner enhancement 
\cite{Feldman:2008xs,Ibe:2008ye,Guo:2009aj} can be achieved 
for the case when $M_{Z'}<2M_\chi$. For this case, the dark 
matter annihilates on the right hand side of the $Z'$ pole. Thus, 
the $\langle \sigma v\rangle_{H}$ in the halo which annihilates near 
the resonance is much larger than $\langle \sigma v\rangle {X_f}$ 
at the  freezeout in the early universe which annihilates far away from the 
resonance. The mass relation also implies that $Z'$ is 
a narrow resonance which is essential for producing sufficient 
amounts of enhancement.

A detailed analysis of the positron excess in StSM is given in Fig.(\ref{fig:pamela}) 
where $M_{Z'}= 298$ GeV, $M_\chi=150$ GeV, and $\epsilon= 0.006$. 
It is shown that the Breit-Wigner enhancement near the $Z'$ pole 
can produce the positron spectrum in the energy range 
$10\sim 100$ GeV observed by PAMELA. We note in passing that 
the Stueckelberg $Z'$ has larger lepton branching ratios than the SM $Z$ 
and other SM like $Z'$ because of the mixing between $U(1)_X$ and 
hypercharge $U(1)_Y$. Thus the $Z'$ does not produce 
too much hadronic final states. 
(For a recent review of different heavy $Z'$ models, see \cite{Langacker:2008yv}.)
The parameter region $M_{Z'}<2M_\chi$ that is favored by 
the PAMELA data is also the region that has promising LHC signatures 
as discussed before.

\section{Conclusion}

The Stueckelberg mechanism allows for an Abelian gauge boson mass growth without 
the benefit of a Higgs mechanism. In the Stueckelberg extension 
of the Standard Model (StSM), the Stueckelberg Lagrangian enlarges 
the neutral vector boson sector which then leads to modifications of the $Z$ boson mass, 
the electric charge, and the hypercharge coupling constant. In the Stueckelberg 
extension of the Standard Model with kinetic mixing (StkSM) but without hidden
sector matter, the kinetic mixing gets absorbed by the Stueckelberg 
mass mixing, which renders StkSM indistinguishable from StSM in the SM electroweak sector. 
A discrimination of the StkSM from StSM comes about due to the presence of 
hidden sector matter. Further, if matter exists in the hidden sector, 
it must be milli charged and here it is interesting 
to note that the milli charge of the hidden sector matter is generated by 
the Stueckelberg mass mixing parameter, and not the kinetic mixing parameter.

The new physics arising in Stueckelberg models is controlled by the Stueckelberg mass mixing 
parameter which must be small to be consistent with the electroweak precision test. However 
this small mixing parameter can still lead to detectable $Z'$ signals which can manifest as 
a narrow resonance at the Tevatron and at the LHC. Tevatron searches from D0 and CDF 
have begun to probe the parameter space of Stueckelberg models in the mass region 
where many other $Z'$s have been excluded. The LHC prediction of the leptonic 
signals of the Stueckelberg $Z'$ shows that one can detect this narrow resonance 
in a broad range of $Z'$ mass.

The milli-charged dark matter in the hidden sector couples with $Z'$ with normal strength, 
so the $Z'$ will decay into the hidden sector predominantly if allowed, and for that case 
the $Z'$ signal is diluted completely at LHC. Due to the weak coupling strength between 
the milli-charged dark matter and the Standard Model particles, the milli-charged dark matter 
has to annihilate in the vicinity of the $Z$ or $Z'$ pole to produce the right amount of 
dark matter to be consistent with the relic abundance measured by WMAP. The PAMELA 
positron excess anomaly can be naturally explained by the Breit-Wigner enhancement 
near the $Z'$ resonance in Stueckelberg models. The Breit-Wigner enhancement 
occurs in the same region of the parameter space where the $Z'$ is a narrow resonance 
and the leptonic signature is strong, and thus detectable at the LHC. \\

\noindent
{\em Acknowledgments}:  
This research was supported in part by NSF grant PHY-0653342.

\end{document}